\documentclass[twocolumn,a4paper,10pt]{article}
\usepackage{amsmath,amsfonts}
\usepackage{epsfig}
\usepackage{times}
\usepackage{epsfig}
\usepackage[small,hang]{caption2}
\usepackage{color}
\usepackage{url}

\definecolor{mygray}{gray}{0.6}
\font\smallfont=cmsy10 at 10truept
\mathchardef\bigCircle="280D

\font\bigfont=cmsy10 at 14.4truept
\mathchardef\tiMes="2902        %

\font\Bigfont=cmsy10 at 17.28truept
\mathchardef\DiaMond="2A05        %
\mathchardef\cirCle="2A0E
\mathchardef\BigCircle="2A0D

\font\Bbigfont=cmsy10 at 24.88truept
\mathchardef\buLLet="2B0F

\def\bigCirc{\raise 0.3ex\hbox{$\bigCircle$}\nobreak$\,$}

\def\Bullet{\raise-0.35ex\hbox{$\buLLet$}\nobreak$\,$}

\def\triangledown{\raise 0.2em\hbox{$\bigtriangledown$}\nobreak$\,$}

\def\minisquare{\hbox{${\vcenter{
               \hrule height 0.3pt \kern-0.4pt
               \hbox{\vrule width  0.3pt height 3.0pt \kern 2.6pt
               \vrule width  0.3pt height 3.0pt} \kern-0.4pt
               \hrule height 0.3pt}}$}}
\def\ssquare{\raise 0.175ex\hbox{${\vcenter{
               \hrule height 0.5truept       \kern-0.25truept
               \hbox{\vrule width 0.5truept height 3.0truept \kern 2.75truept
                     \vrule width 0.5truept height 3.0truept} \kern-0.25truept
               \hrule height 0.5truept}}$}\nobreak$\,$}
\def\squarex{\raise 0.175ex\hbox{${\vcenter{
               \hrule height 0.8truept       \kern-1.80truept
          \hbox{\vrule width 0.8truept height 8.0truept \kern-1.95truept
                \raise 0.8truept\hbox{$\tiMes$}     \kern-6.70truept
                \vrule width 0.8truept height 8.0truept} \kern-0.80truept
               \hrule height 0.8truept}}$}\nobreak$\,$}

\def\sqbull{\raise0.175ex\hbox{\vrule height 1.4ex width 1.6ex depth 0.2ex}\nobreak$\,$}
\def\smsqbull{\raise0.175ex\hbox{\vrule height 0.8ex width 0.9ex depth 0.2ex}\nobreak$\,$}

\def\Diamondplus{${\vcenter{\vcenter{\DiaMond} \kern-10truept
                            \hbox{\vrule width .4truept}\kern -3truept
                            \hrule height .4truept}}$\nobreak$\,$}
\newcount\ndots

\def\drawline#1#2{\raise 2.5truept\vbox{\hrule width #1truept height #2truept}}
\def\moonspace#1{\hskip #1truept}

\def\Dashy{\drawline{4.00}{1.00}}     
\def\dashy{\drawline{4.00}{0.75}}     
\def\thindashy{\drawline{4.00}{0.25}}     
\def\dashyspace{\dashy\moonspace{2}}
\def\Dashyspace{\Dashy\moonspace{2}}
\def\thindashyspace{\thindashy\moonspace{2}}
\def\longdashy{\drawline{8.00}{0.75}} 
\def\thinlongdashy{\drawline{8.00}{0.25}} 
\def\longdashyspace{\longdashy\moonspace{2}}
\def\thinlongdashyspace{\thinlongdashy\moonspace{2}}

\def\solid{\drawline{24}{0.75}\nobreak$\,$}

\def\dashbox{\hbox{\dashyspace}}  
\def\Dashbox{\hbox{\Dashyspace}}  
\def\dashed{\hbox {\ndots=0 \loop\ifnum\ndots<3\advance\ndots by 1
        \dashbox\repeat\dashy}\nobreak$\,$}       
\def\Dashed{\hbox {\ndots=0 \loop\ifnum\ndots<3\advance\ndots by 1
        \Dashbox\repeat\Dashy}\nobreak$\,$}       
\def\thindashbox{\hbox{\thindashyspace}}  
\def\thindashed{\hbox {\ndots=0 \loop\ifnum\ndots<3\advance\ndots by 1
        \thindashbox\repeat\thindashy}\nobreak$\,$}       
\def\thindash{\hbox {\ndots=0 \loop\ifnum\ndots<3\advance\ndots by 1
        \thindashbox\repeat\thindashy}\nobreak$\,$}       

\def\longdashbox{\hbox{\longdashyspace}}  
\def\thinlongdashbox{\hbox{\thinlongdashyspace}}  
\def\longdash{\hbox {\ndots=0 \loop\ifnum\ndots<3\advance\ndots by 1
        \longdashbox\repeat\longdashy}\nobreak$\,$}       
\def\thinlongdash{\hbox {\ndots=0 \loop\ifnum\ndots<3\advance\ndots by 1
        \thinlongdashbox\repeat\thinlongdashy}\nobreak$\,$}       

\makeatletter

\newcounter{numbersec}
\renewcommand{\section}[1]{\par\noindent\stepcounter{numbersec}
\par
\vspace{6pt}
\noindent\textbf{\large   \arabic{numbersec} \hspace*{0.3cm} #1 }
\par
\vspace{2pt}
}
\renewcommand{\subsection}[1]{
\par
\vspace{6pt}
\noindent\textbf{#1}
\par
}
\renewcommand{\subsubsection}[1]{%
\par
\vspace{6pt}
\textbf{#1.}
}

\newcommand{\Abstract}{\par\vspace{6pt}\noindent\textbf{\large Abstract}\par\vspace{2pt}}
\newcommand{\Acknowledgments}{\par\vspace{6pt}\noindent\textbf{\large Acknowledgments }\par\vspace{2pt}}

\newenvironment{References}{
\par\vspace{6pt}\noindent\textbf{\large References}\par\vspace{2pt}
\begin{small}\begin{list}{ }{
\itemsep0mm \parsep0mm\labelsep0mm\leftmargin0mm
}}
{\end{list}\end{small}}

\makeatother

\title{\vspace*{-12mm}
\LARGE \sc \textbf{  
Assessment of turbulent boundary layers on a \\ NACA4412 wing section
at moderate \boldmath{$Re$}
}}
\author{ \Large \bf \textit{ 
R. Vinuesa$^{1}$, S. M. Hosseini$^{1}$,} \\
\Large \bf \textit{A. Hanifi$^{1,2}$, D. S. Henningson$^{1}$ and P. Schlatter$^{1}$}  \\ \\
\bf  $^{1}$ \textit{Linn\'e FLOW Centre, KTH Mechanics, SE-100 44 Stockholm, Sweden} \\
\bf \textit{and Swedish e-Science Research Centre (SeRC), Stockholm, Sweden}\\
\bf  $^{2}$ \textit{Swedish Defence Research Agency, FOI, Stockholm, Sweden} \\ \\
\underline{\bf rvinuesa@mech.kth.se}
}
\date{}

\begin{document}
%


%

\maketitle
\thispagestyle{empty}



%
%
\Abstract
The results of a DNS of the flow around a wing section represented by a NACA4412 profile, with $Re_{c} = 400,000$ and $5^{\circ}$ angle of attack, are presented in this study. The high-order spectral element code Nek5000 was used for the computations. The Clauser pressure-gradient parameter $\beta$ ranges from $\simeq 0$ and 85 on the suction side, and the maximum $Re_{\theta}$ and $Re_{\tau}$ values are around $2,800$ and 373, respectively. Comparisons between the suction side with ZPG TBL data show a more prominent wake, a steeper logarithmic region and lower velocities in the buffer region. The APG also leads to a progressively increasing value of the inner peak in the tangential velocity fluctuations, as well as the development of an outer peak, which is also observed in the other components of the Reynolds stress tensor. Other effects of strong APGs are increased production and dissipation profiles across the boundary layer, together with enhanced viscous diffusion and velocity-pressure-gradient correlation values near the wall. All these effects are connected to the fact that the large-scale motions of the flow become energized due to the APG, as apparent from spanwise premultiplied power spectral density plots.

\section{Introduction}

Despite their great technological importance, the turbulent boundary layers developing around wing sections have not been characterized in detail in the available literature. One of the most remarkable studies in this regard is the work by Coles (1956) 60 years ago, where he, among other aspects, analyzed several sets of measurements on airfoils approaching separation, and he introduced the concept of the ``law of the wake''. Progressive increase in computer power has allowed in the recent years to perform numerical simulations on relatively complex geometries, which have shed some light on the physics taking place on wing sections. Some examples are the direct numerical simulations (DNSs) of Jones {\it et al.} (2008) and the large-eddy simulations (LESs) of Alferez {\it et al.} (2013), at Reynolds numbers based on freestream velocity $U_{\infty}$ and wing chord length $c$ of $Re_{c}=50,000$ and 100,000, respectively. Nevertheless, these studies focus on laminar separation bubbles (LSBs), and therefore do not allow to characterize the development of the turbulent boundary layers throughout the suction and pressure sides of the wing (which will be denoted as $ss$ and $ps$, respectively). 

In the present study we report the results of a DNS of the flow around a NACA4412 wing section, at an unprecedented $Re_{c}=400,000$, with $5^{\circ}$ angle of attack. Although incipient separation is observed beyond $x_{ss}/c \simeq 0.9$ ($x$ being the chord-wise coordinate), the mean skin friction coefficient $C_{f}$ is always positive, which indicates that the mean flow is attached throughout the whole wing. Note that $C_{f}=2 \left ( u_{\tau} / U_{e} \right )^{2}$, where $U_{e}$ is the local velocity at the boundary-layer edge and $u_{\tau}$ is the friction velocity. Therefore the emphasis of this work is on the streamwise development of the turbulent boundary layers developing around the wing, and the effect of the pressure gradient on the most relevant turbulent features. 

\section{Numerical method}

\begin{figure*}[ht]
\begin{center}
\includegraphics[width=1\linewidth]{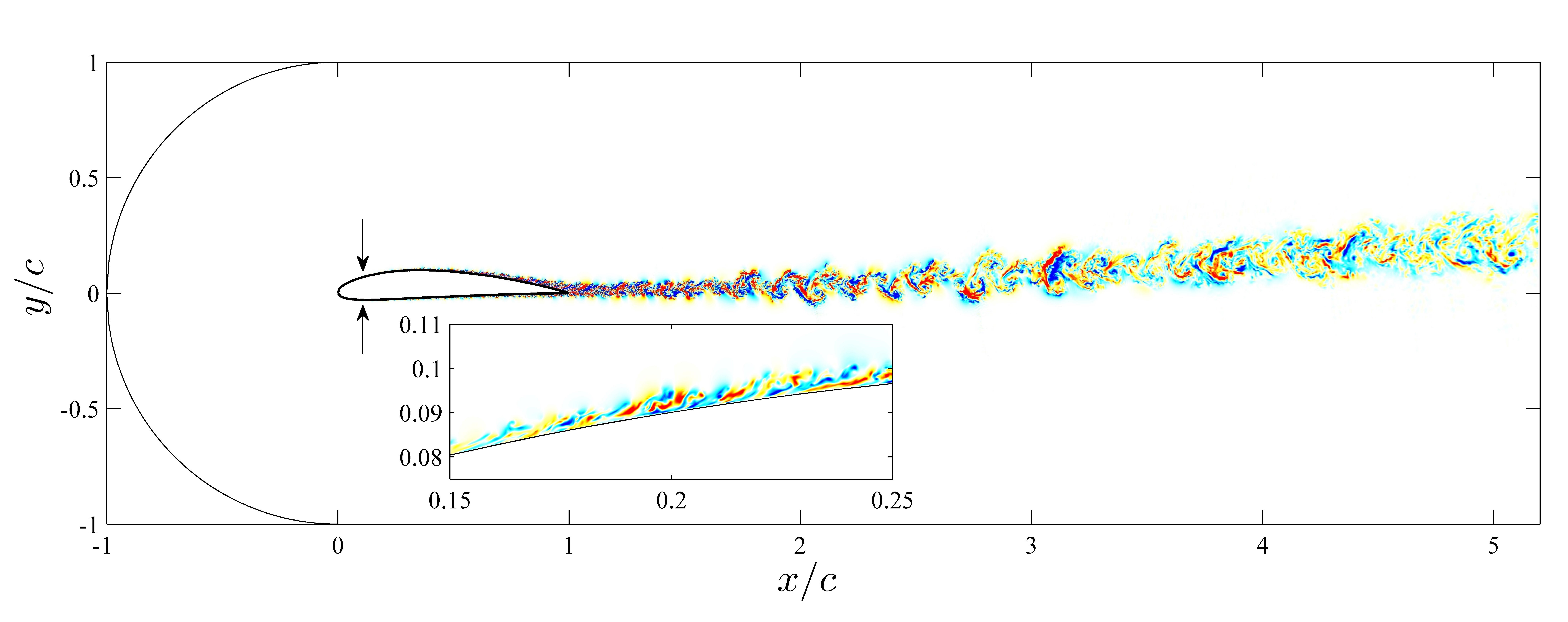}
\caption{\label{visualization_figure} Two-dimensional slice of the computational domain showing with arrows the locations where the flow is tripped. Instantaneous spanwise velocity is also shown, where blue and red indicate positive and negative values, respectively. The insert shows a detailed view of the flow on the suction side of the wing, and the spanwise velocities range from $-0.52$ to $0.52$.}   
\end{center}
\end{figure*}

In order to properly simulate the complex multi-scale character of turbulence, it is essential to use high-order numerical methods. The DNS described in this work was carried out with the code Nek5000 (Fischer {\it et al.}, 2008), which is based on the spectral element method, and Lagrange interpolants of polynomial order $N=11$ were considered for the spatial discretization. The computational domain has chord-wise and vertical lengths $L_{x}=6.2c$ and $L_{y}=2c$ respectively, and the periodic spanwise direction has a length of $L_{z}=0.1c$. As can be observed in Figure \ref{visualization_figure}, we considered a C-mesh; a Dirichlet boundary condition extracted from a previous RANS simulation was imposed in all the boundaries except at the outflow, where the natural stress-free condition was used. A total of 1.85 million spectral elements was employed to discretize the domain, which amounts to around 3.2 billion grid points. Moreover, the boundary layers developing over the suction and pressure sides of the wing were tripped using the volume-force approach proposed by Schlatter and \"Orl\"u (2012), at a chord-wise distance of $x/c=0.1$ from the wing leading edge. The mesh was designed in order to satisfy the condition $h \equiv \left ( \Delta x \cdot \Delta y \cdot \Delta z \right )^{1/3} < 5 \eta$ everywhere in the domain, where $\eta= \left ( \nu^{3} / \varepsilon \right)^{1/4}$ is the Kolmogorov scale and $\varepsilon$ is the local isotropic dissipation, so that the mesh is fine enough to capture the smallest turbulent scales. This can be observed in the level of detail obtained even in the near-wall region in Figure \ref{visualization_figure}. A comprehensive description of the setup can be found in the work by Hosseini {\it et al.} (2016).

\section{Turbulence statistics} \label{statistics_section}

\begin{figure*}[ht]
\begin{center}
\includegraphics*[width=0.3\linewidth]{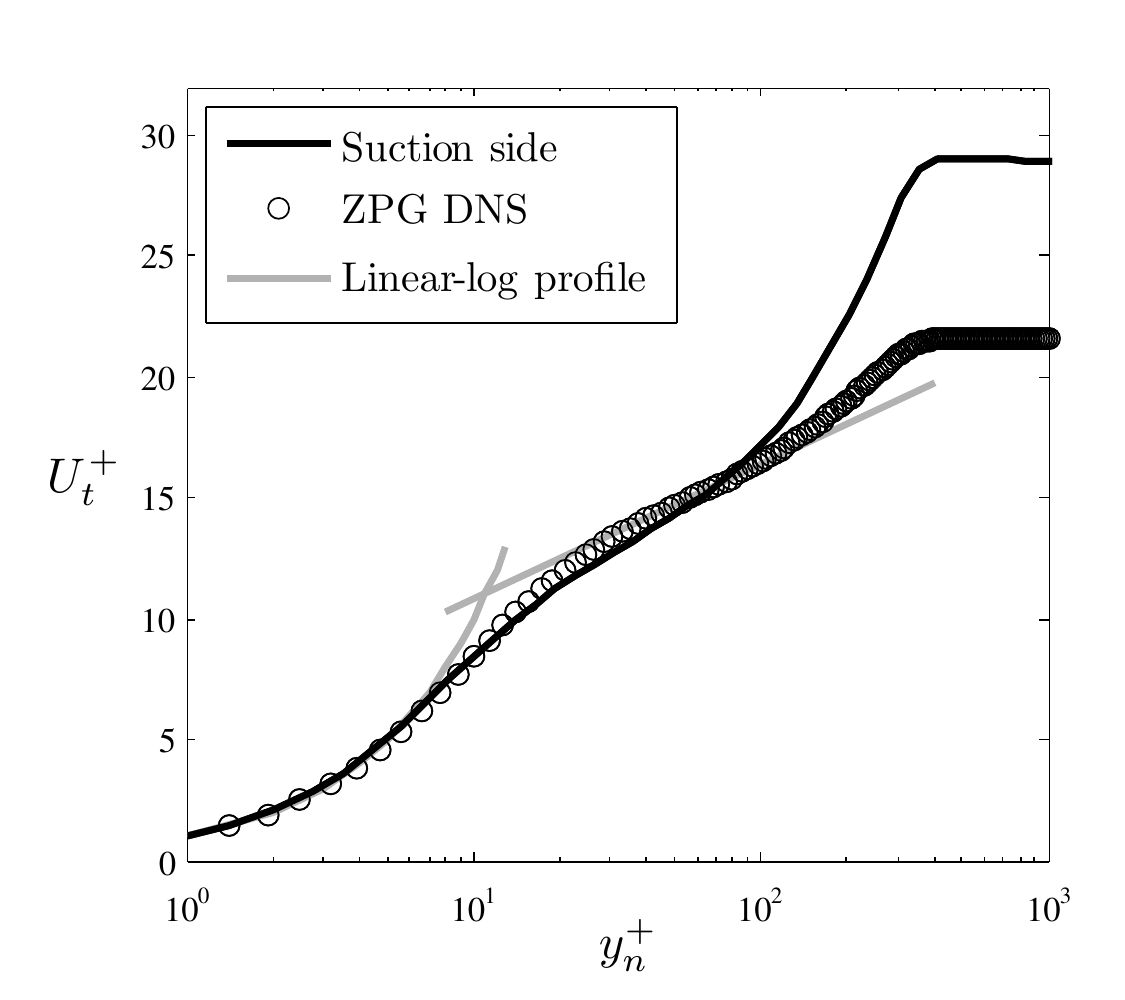}
\includegraphics*[width=0.3\linewidth]{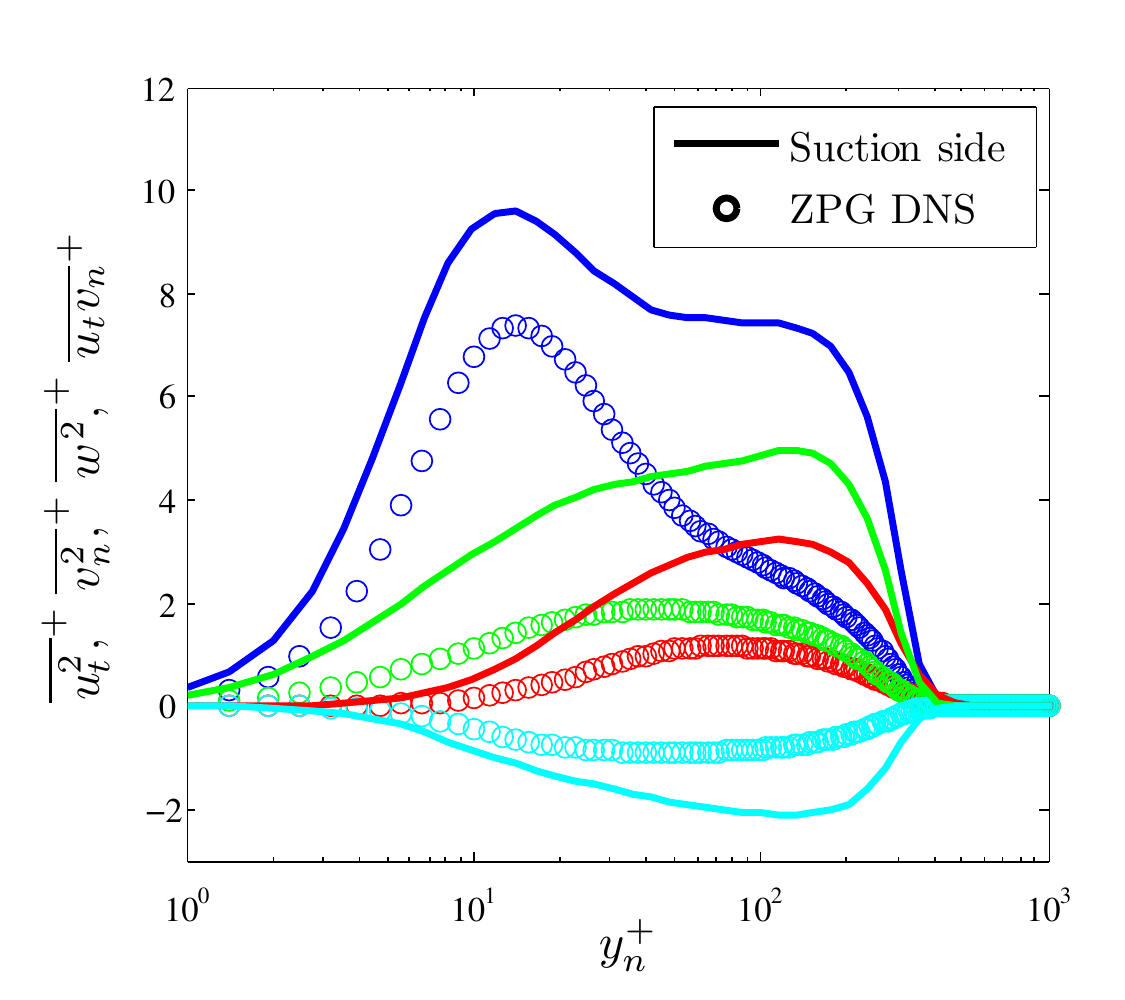}
\includegraphics*[width=0.3\linewidth]{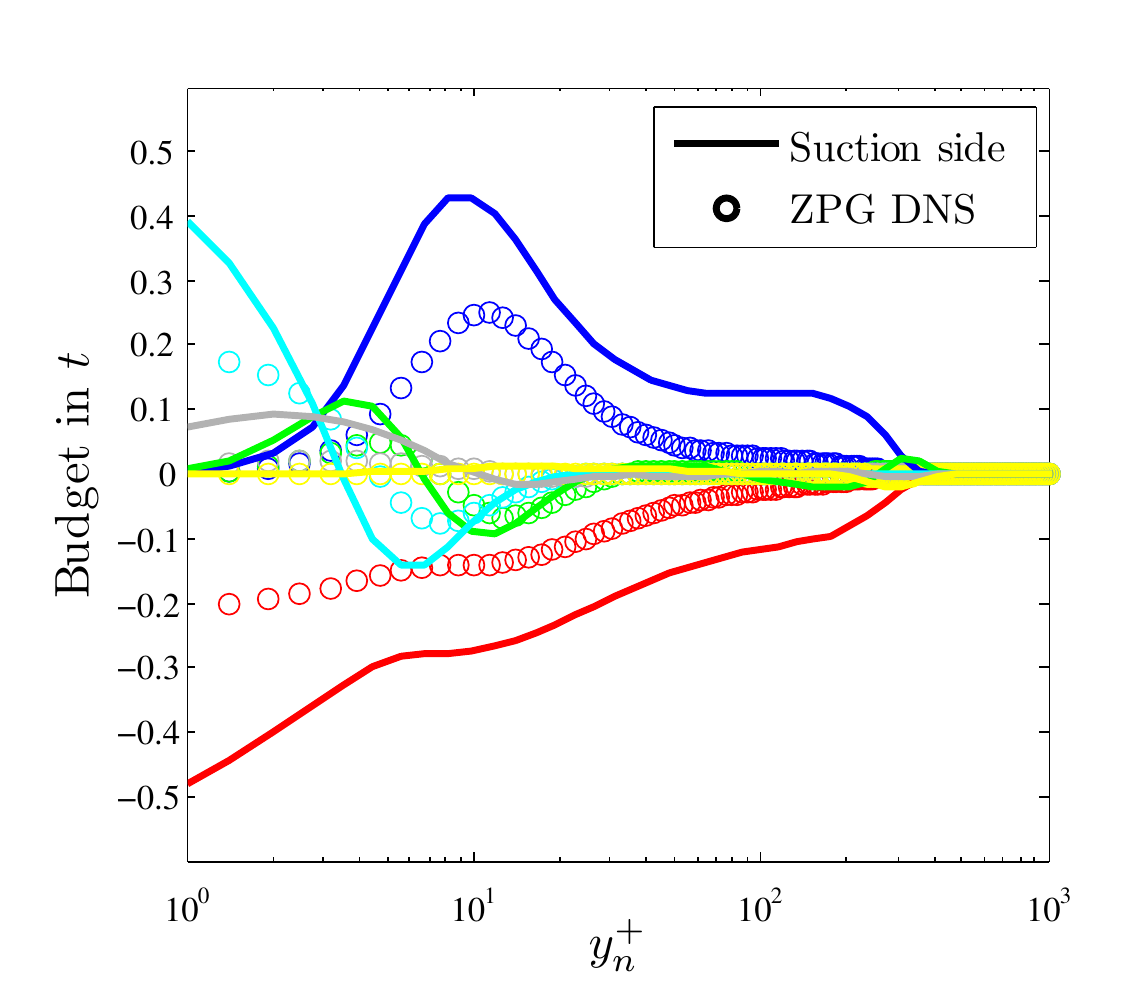} \\
\includegraphics*[width=0.3\linewidth]{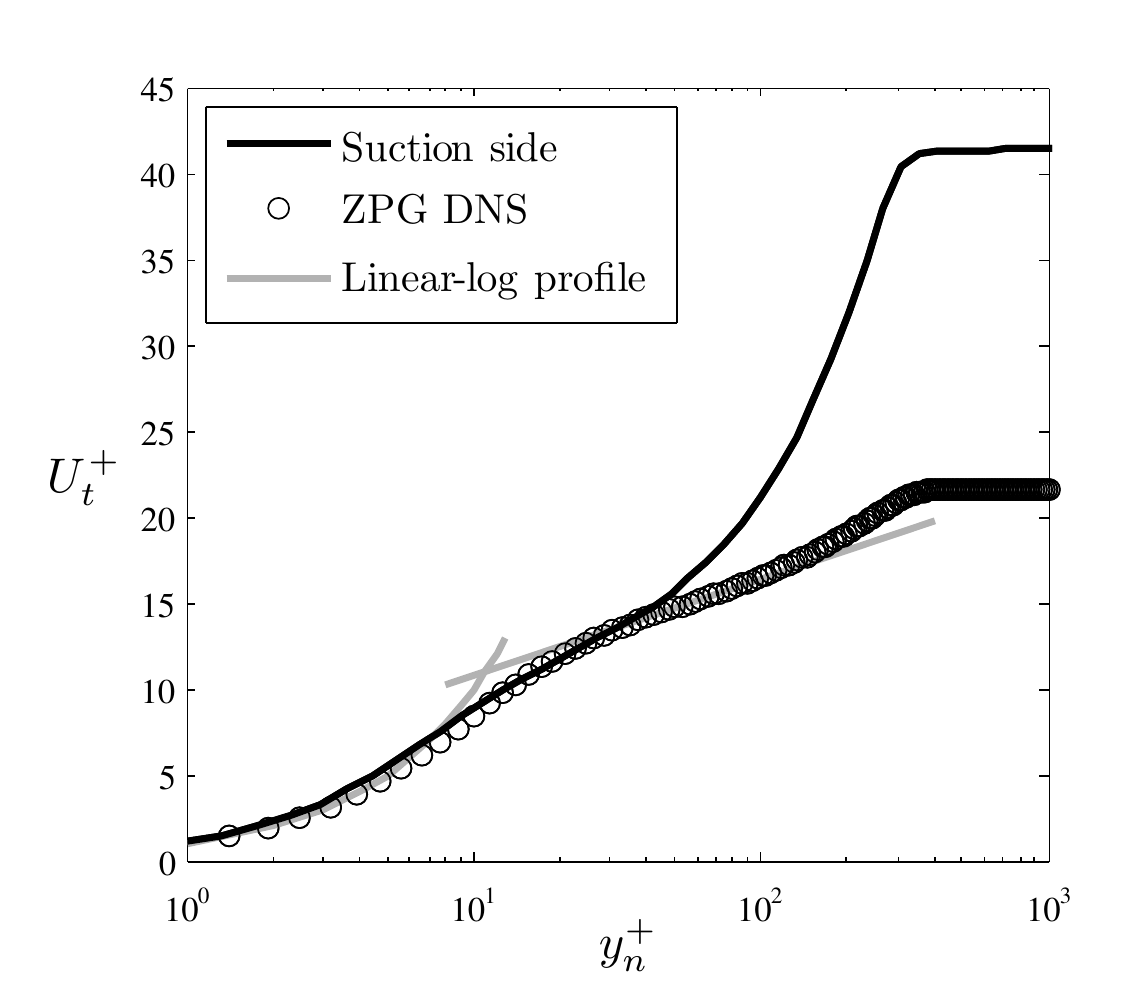}
\includegraphics*[width=0.3\linewidth]{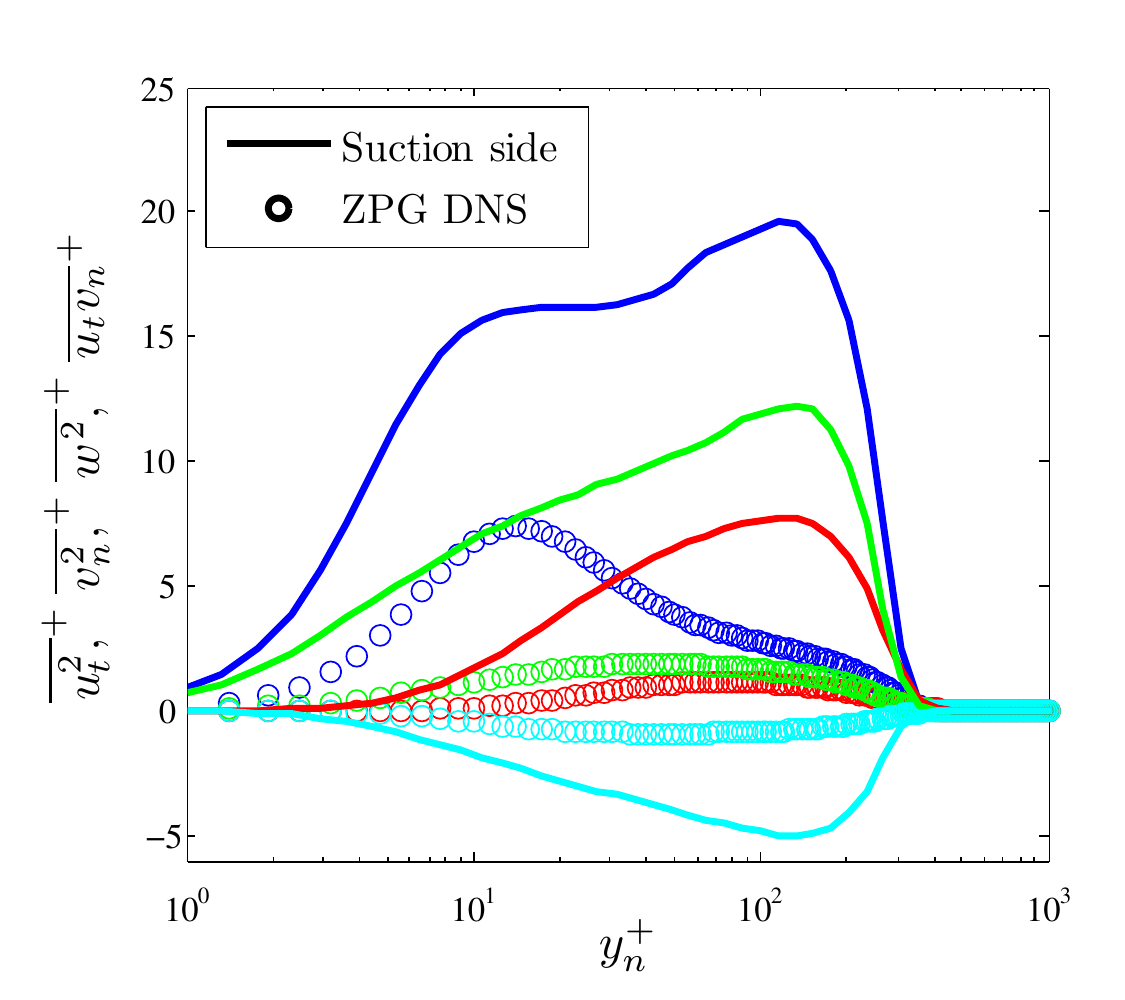}
\includegraphics*[width=0.3\linewidth]{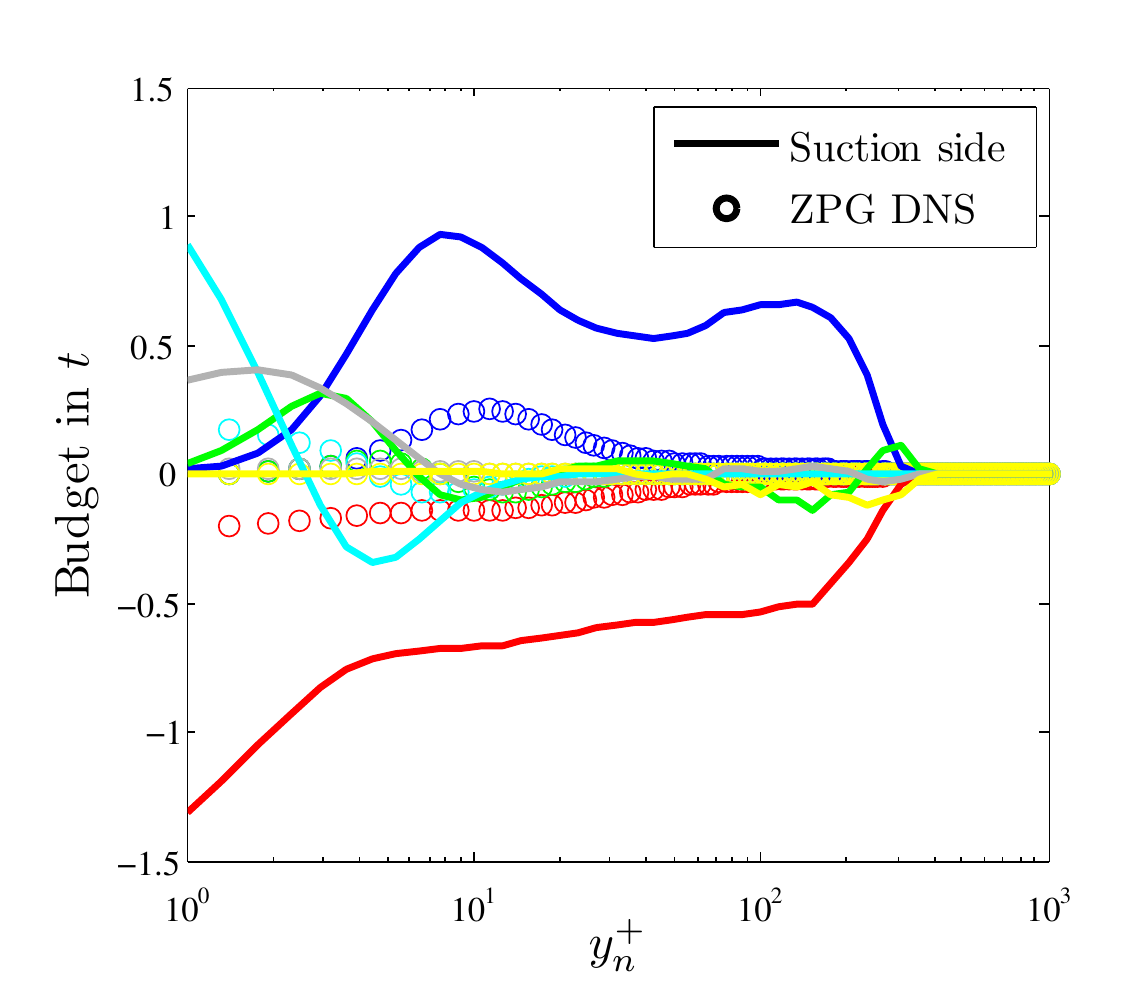} \\
\caption{\label{statistics_figure} (Left panel) Inner-scaled mean flow (with reference low-$Re$ values $\kappa=0.41$ and $B=5.2$), (middle panel) inner-scaled Reynolds-stress tensor components and (right panel) TKE budget scaled by $u_{\tau}^{4}/\nu$. Reynolds stresses are represented as: {\color{blue}\solid} tangential, {\color{red}\solid} wall-normal and {\color{green}\solid} spanwise velocity fluctuations, and {\color{cyan}\solid} Reynolds shear stress. Budget terms are represented as follows: {\color{blue}\solid}  Production, {\color{red}\solid} Dissipation, {\color{green}\solid} Turbulent transport, {\color{cyan}\solid} Viscous diffusion, {\color{mygray}\solid} Velocity-pressure-gradient correlation and {\color{yellow}\solid} Convection. Data extracted at (top line) $x_{ss}/c = 0.8$ and (bottom line) $x_{ss}/c = 0.9$, and compared with the ZPG data by Schlatter and \"Orl\"u (2010).}   
\end{center}
\end{figure*}

In order to compute complete turbulence statistics, the simulation was run for a total of 10 flow-over times, which correspond to at least 12 eddy-turnover times (defined as $\delta_{99}/u_{\tau}$, where $\delta_{99}$ is the $99\%$ boundary layer thickness) throughout the whole wing except for $x_{ss}/c > 0.9$. Note that this region is subjected to a very strong adverse pressure gradient (APG), and therefore the turbulent scales are significantly larger than in the rest of the wing. The boundary layers developing around the wing were characterized at a total of 80 profiles on both sides, projected on the directions tangential ($t$) and normal ($n$) to the wing surface, and the magnitude of the pressure gradient was quantified in terms of the Clauser pressure-gradient parameter $\beta= \delta^{*} / \tau_{w} {\rm d}P_{e} / {\rm d} x_{t}$. Note that $\delta^{*}$ is the displacement thickness, $P_{e}$ is the pressure at the boundary-layer edge and $x_{t}$ is the coordinate tangential to the wing surface. Figure \ref{statistics_figure} shows mean flow, Reynolds-stress tensor components and turbulent kinetic energy (TKE) budgets at $x_{ss}/c = 0.8$ and 0.9, respectively. The boundary layer is subjected to a strong APG at $x_{ss}/c = 0.8$, where the value of $\beta$ is 4.1, and as can be observed in Figure \ref{statistics_figure} (top) the APG leads to a more prominent wake region (as also reported by Monty {\it et al.} (2011) and Vinuesa {\it et al.} (2014)), a steeper incipient log region, and reduced velocities in the buffer layer compared with the DNS of ZPG boundary layer by Schlatter and \"Orl\"u (2010). Table \ref{x_0809t} shows several mean flow parameters of the boundary layer at $x_{ss}/c = 0.8$ compared with the ZPG at approximately matching friction Reynolds number $Re_{\tau}=\delta_{99} u_{\tau} / \nu$ (where $\nu$ is the fluid kinematic viscosity). Note that the difficulties of determining the boundary-layer thickness in pressure-gradient TBLs were discussed by Vinuesa {\it et al.} (2016), and their method was considered in the present study to calculate $\delta_{99}$. The APG effectively lifts up the boundary layer and increases its thickness, which leads to a larger shape factor $H=\delta^{*} / \theta$ (where $\theta$ is the momentum thickness), and also to a reduced skin friction coefficient. The lower value of the von K\'arm\'an coefficient $\kappa$ is connected with a steeper log law, and the larger wake parameter $\Pi$ shows the strong impact on the wake region. As shown by Monty {\it et al.} (2011), the APG energizes the large-scale structures in the flow, which have a strong interaction with the outer flow (thus the impact on the wake region). These large-scale motions are usually wall-attached eddies, which leave their footprint at the wall and therefore significantly affect the overlap and buffer layers. Additional insight on the effect of pressure gradients on the turbulent boundary layers developing around the wing can be achieved by analyzing the components of the Reynolds-stress tensor also shown in Figure \ref{statistics_figure}. The impact of the APG can clearly be observed at $x_{ss}/c=0.8$ on the tangential velocity fluctuations $\overline{u^{2}_{t}}^{+}$: the inner peak is increased, and the effect on the outer region is quite noticeable, as also observed by Sk\r{a}re and Krogstad (1994), Marusic and Perry (1995) and Monty {\it et al.} (2011). This is associated with the largest and most energetic scales in the flow interacting with the APG, as is also noticeable from the larger values of $\overline{w^{2}}^{+}$ in the outer region. Note that the tangential velocity fluctuation profile starts to develop an outer peak, as also observed by Monty {\it et al.} (2011), which is connected to the fact that the structures in the outer flow are more energetic due to the effect of the APG. It is also interesting to note that the effect on the wall-normal velocity fluctuations $\overline{v^{2}_{n}}^{+}$ and the Reynolds shear stress is also significant, although slightly less pronounced. Figure \ref{statistics_figure} (top) also shows the TKE budget at $x_{ss}/c=0.8$, and it is interesting to note that the effect of the pressure gradient is noticeable in all the terms.  More specifically, the APG leads to an increased inner peak in the production profile (around $70\%$ larger than the one in the ZPG boundary layer), which is connected to the increased peak in tangential velocity fluctuations, as well as an incipient peak in the outer region. The effect on the dissipation is significant in the near-wall region, which shows enhanced dissipation levels (around $90\%$ larger than the ZPG TBL), although the discrepancy with respect to the ZPG case progressively diminishes as the outer region is approached. Interestingly, the viscous diffusion is also increased in the near-wall region as a consequence of the APG, and when it becomes negative it also exhibits larger values than the ZPG TBL, in this case to balance the rapidly growing production. Beyond $y^{+}_{n} \simeq 10$ the APG profile converges to the one from the ZPG. Therefore the interactions between the large-scale motions in the outer region have a manifestation in the redistribution of TKE terms close to the wall, as can also be observed in the increased values of the velocity-pressure-gradient correlation for $y^{+}_{n} < 10$, which is positive, and also balances the increased dissipation. 

The TBL on the suction side of the wing is subjected to a strong APG with $\beta \simeq 14.1$ at $x_{ss}/c = 0.9$, and its statistics are shown in Figure \ref{statistics_figure} (bottom). The inner-scaled mean flow is shown on the left panel of Figure \ref{statistics_figure}, and the most relevant flow parameters are summarized in Table \ref{x_0809t}. The effects observed in the moderate-APG case at $x_{ss}/c = 0.8$ are even more noticeable in this case, where the impact on the wake parameter, incipient log layer and buffer region is even larger. Similar effects on the mean flow can be observed in the experimental study by Sk\r{a}re and Krogstad (1994) with a comparably large $\beta$ value of 19.9, at $Re_{\theta}$ up to 39,120. 
Regarding the Reynolds-stress tensor components, the first interesting observation is the fact hat the inner peak in the tangential velocity fluctuation profile exceeds the one from the ZPG by a factor of around 2, and the outer peak is around $33\%$ larger than the inner one. The other components of the Reynolds-stress tensor also exhibit significantly larger values in the outer region compared with the ZPG case, which again shows the effect of the APG energizing the large-scale motions of the flow, and in particular the significantly modified Reynolds shear stress shows the very different momentum distribution mechanisms across the boundary layer under the effect of the APG. Although Sk\r{a}re and Krogstad (1994) did not take measurements close to the wall, they also characterized the significantly large peaks in the outer region of the various components of Reynolds stress tensor. In this sense, it can be argued that APG TBLs exhibit features of higher Reynolds number boundary layers,  as also pointed out by Harun {\it et al.} (2013), who compared the features of TBLs subjected to APG, ZPG and FPG conditions, and suggested the possibility of connecting high $Re$ effects in ZPG boundary layers with the effect of APGs.
In this context, Hutchins and Marusic (2007) showed how the energy of the turbulent structures in the log region increases with $Re$, becoming comparable with the energy in the near-wall region. This was also observed in the experiments by Vallikivi {\it et al.} (2015) on pressurized ZPG boundary layers up to $Re_{\theta} \simeq 223 \times 10^{3}$, which start to exhibit a prominent outer peak in the streamwise velocity fluctuation profile, of magnitude comparable to the one of the inner peak. However, a proper assessment of these effects would require investigations of numerical and experimental nature at much higher Reynolds numbers, in order to properly isolate Reynolds number and pressure gradient effects. Regarding the TKE budget on the right panel, both production and dissipation profiles exceed by at least a factor of 4 the ones of the ZPG throughout the whole boundary layer. It is also remarkable the emergence of an outer peak in the production profile, which is around $40\%$ lower than the inner production peak. This phenomenon was also observed by Sk\r{a}re and Krogstad (1994) in their experimental boundary layer with $\beta \simeq 19.9$ and $Re_{\theta} \simeq 39,120$, although in their case the magnitude of the outer peak was almost as large as the one from the inner peak, and they found it farther away from the wall: at $y/\delta \simeq 0.45$, whereas in our case it is located at $y/\delta_{99} \simeq 0.35$. It can be argued that the discrepancy in magnitude and location of this outer peak is caused both by APG strength and Reynolds-number effects. Sk\r{a}re and Krogstad (1994) also showed that there was considerable diffusion of turbulent kinetic energy from the central part of the boundary layer towards the wall, which was produced by the emergence of this outer peak. Since in our case the outer peak of the streamwise velocity fluctuations is larger than the inner peak, but in the production profile the outer peak is smaller, it could be conjectured that the APG effectively energizes the large-scale motions of the flow, and eventually these more energetic structures become a part of the production mechanisms characteristic of wall-bounded turbulence. The high levels of dissipation observed in our case also far from the wall were also reported in the experiment by Sk\r{a}re and Krogstad (1994), and in particular they also documented the presence of the inflection point in the dissipation profile at roughly the same wall-normal location as the outer peak of the production. Other relevant terms significantly affected by the APG are the viscous diffusion, which again shows larger values very close to the wall to balance the increased dissipation, and in this case changes sign at an even lower value of $y^{+}_{n}$:  $\simeq 2.5$. The velocity-pressure-gradient correlation also shows significantly increased values close to the wall compared with the ZPG case, but as in the $\beta \simeq 4.1$ APG, for $y^{+}_{n} > 10$ both the viscous diffusion and the velocity-pressure-gradient profiles approximately converge to the ZPG ones. In addition to the increased maxima of turbulent transport and convection observed close to the boundary-layer edge, this strong APG case exhibits a relative minimum of turbulent transport at approximately the same location as the outer production peak, which is interesting because beyond this location this term changes sign. This suggests that the very strong production in the outer region leads to additional negative turbulent transport to balance, together with the dissipation, this locally increased production level.
\begin{table}
\begin{center}
\def~{\hphantom{0}}
\begin{tabular}{c c c c}
Parameter & $x_{ss}/c = 0.8$ & $x_{ss}/c = 0.9$ & ZPG DNS \\[3pt]
\hline
$Re_{\tau} $ & 373 & 328 &	359 \\
$\beta$	  & 4.1 & 14.1&	$\simeq 0$ \\
$Re_{\theta}$ & 1,722 &  2,255 &	1,007 \\
$H$	 & 1.74 & 2.03&	1.45	\\
$C_{f} $& $2.4 \times 10^{-3}$ & $ 1.2 \times 10^{-3}$ &	$4.3 \times 10^{-3}$	\\
$ \kappa $ & 0.33 & 0.23 &	0.41 \\
$B $ & 2.08	& -2.12 & 4.87 \\
$ \Pi $ & 1.35 & 1.83 &	0.37 \\
\end{tabular}
\caption{Boundary-layer parameters at $x_{ss}/c = 0.8$ and 0.9, compared with ZPG results by Schlatter and \"Orl\"u (2010).}   
\label{x_0809t}
\end{center}
\end{table}

\section{Spectral analysis}

\begin{figure}[ht]
\begin{center}
\includegraphics*[width=0.95\linewidth]{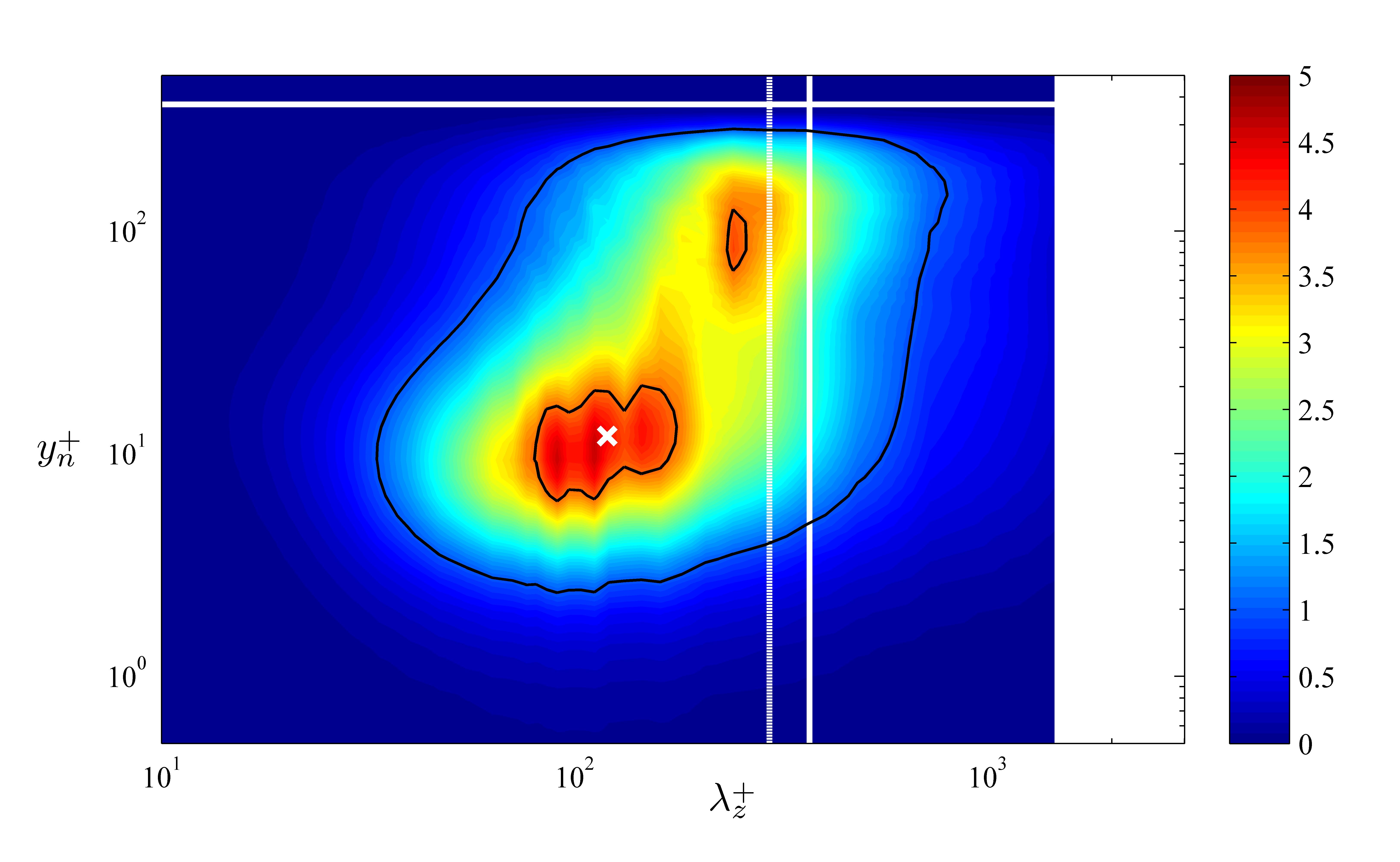}
\includegraphics*[width=0.95\linewidth]{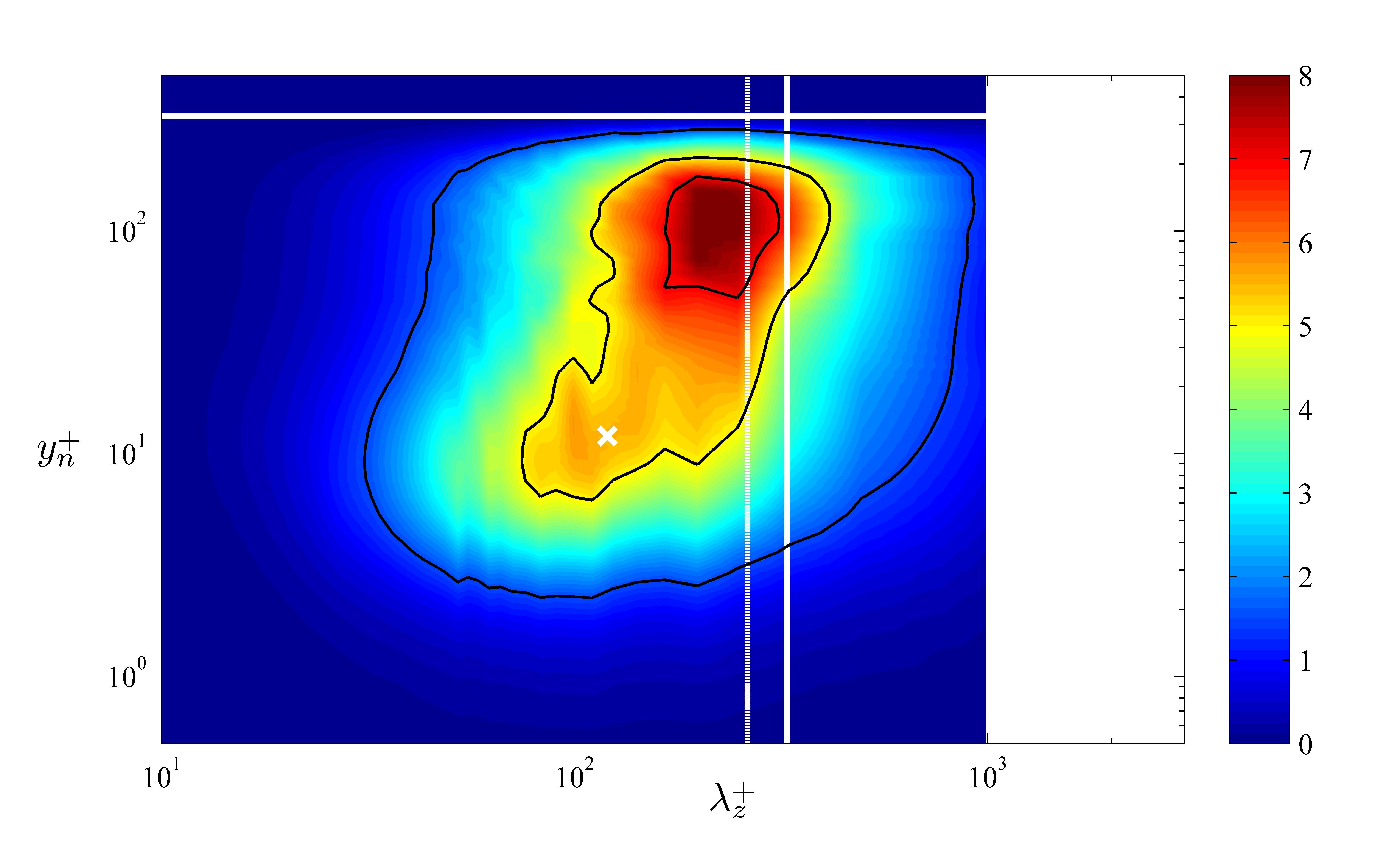}
\caption{\label{spectra_figure} Inner-scaled spanwise premultiplied power spectral density of the tangential velocity $k_{z} \Phi _{u_{t} u_{t}} / u_{\tau}^{2}$. Spectra calculated at (top) $x_{ss}/c = 0.8$ and (bottom) $x_{ss}/c = 0.9$. White crosses indicate the location $y^{+}_{n}=12$, $\lambda_{z}^{+}=120$, white solid lines denote the inner-scaled boundary layer thickness $\delta_{99}^{+}$, and white dashed lines show the position $\lambda_{z} \simeq 0.8 \delta_{99}$. Black solid lines indicate contour levels of 1 and 3.8 at $x_{ss}/c = 0.8$, and 1.5, 5 and 7 at $x_{ss}/c = 0.9$.}   
\end{center}
\end{figure}

In order to further assess the characteristics of the boundary layers developing around the wing section, their energy distribution is studied through the analysis of the inner-scaled spanwise premultiplied power spectral density of the tangential velocity $k_{z} \Phi _{u_{t} u_{t}}^{+}$, shown at $x_{ss}/c=0.8$ and 0.9 in Figure \ref{spectra_figure}. The first interesting feature of these spectra is the fact that they exhibit the so-called inner-peak of spectral density, at a wall-normal distance of around $y_{n}^{+} \simeq 12$, and for wavelenghts of around $\lambda_{z}^{+} \simeq 120$. This was also observed in the LES of ZPG boundary layer by Eitel-Amor {\it et al.} (2014) up to a much higher $Re_{\theta}=8,300$, and is a manifestation of the inner peak of the tangential velocity fluctuations discussed in $\S$3. In fact, the value of this inner peak is also highly affected by the pressure gradient: at $x_{ss}/c=0.8$ it takes a value of around 5, which is larger than the value of approximately 4 in ZPG TBLs, and at $x_{ss}/c=0.9$ it rises up to 6. This behavior strongly resembles the one of the tangential velocity fluctuations, and highlights the connection between the coherent structures in the boundary layer and the turbulence statistics. Moreover, the wavelength $\lambda_{z}^{+} \simeq 120$ corresponds to the characteristic streak spacing in wall-bounded turbulence, as shown for instance by Lin {\it et al.} (2008). In this context, it is also interesting to note that the domain is sufficiently wide to capture the contributions of all the relevant turbulent scales in the boundary layer, even in the strongly decelerated and very thick boundary layer conditions found at $x_{ss}/c=0.9$.

Regarding the spectra in the outer region of the boundary layer, it is first interesting to note the emergence of an outer peak with a value of inner-scaled power spectral density of around 4 at $x_{ss}/c=0.8$. The very strong APG found at $x_{ss}/c=0.9$  leads to a power spectral density level on the outer region larger than the one in the inner region of the boundary layer, with an inner-scaled value of around 8. The connection with the streamwise turbulence intensity profiles is again clear in the development of the outer region, since at $x_{ss}/c=0.8$ the outer peak is also slightly below the inner one (but of the same magnitude as the inner peak in a ZPG boundary layer), and at $x_{ss}/c=0.9$ also in the $\overline{u^{2}_{t}}^{+}$ profile the outer peak is larger than the inner one. Therefore, the progressively stronger APG energizes the large-scale motions of the flow, which on the other hand have a footprint in the near-wall region responsible for the increase of energy in the buffer layer. The emergence of this outer spectral peak was also observed by Eitel-Amor {\it et al.} (2014) in their ZPG simulations at much higher Reynolds numbers, with an incipient outer peak at $Re_{\theta} \simeq 4,400$ which started to become more noticeable at around $Re_{\theta} \simeq 8,300$. Note that in their case the spectral density level in the outer region was significantly lower than the one in the inner region, and therefore much higher Reynolds numbers would be necessary in a ZPG boundary layer in order to reach similar outer energy levels. On the other hand, Eitel-Amor {\it et al.} (2014) observed the emergence of the outer spectral peak at around $\lambda_{z} \simeq 0.8 \delta_{99}$, whereas the results in Figure \ref{spectra_figure} show that in the suction side of the wing the outer peak emerges at around $\lambda_{z} \simeq 0.65 \delta_{99}$. Due to the significantly lower Reynolds numbers present in the wing, it is difficult to assess whether this difference in the structure of the outer region is due to a fundamentally different mechanism in the energizing process of the large-scale motions from APGs and high-$Re$ ZPGs, or whether this is due to low-$Re$ effects. In any case, and as also noted by Harun {\it et al.} (2013), the effect of the pressure gradient on the large-scale motions in the flow has features in common with the effect of $Re$ in ZPG boundary layers, and therefore further investigation at higher Reynolds numbers would be required to separate pressure-gradient and Reynolds-number effects.

\section{Conclusions}

In the present study we perform a DNS of the flow around a wing section represented by a NACA4412 profile, with $Re_{c}=400,000$ and $5^{\circ}$ angle of attack. The high-order spectral element code Nek5000 is used for the computations, which are carried out with 16,384 cores on the Cray XC40 system ``Beskow'' at KTH, Stockholm.  The Clauser pressure-gradient parameter $\beta$ ranges from $\simeq 0$ and 85 on the suction side, and thus this TBL is subjected to a progressively stronger APG.  The first effect of the APG on the mean flow is the more prominent wake, reflected in a larger $U^{+}_{e}$ and a larger wake parameter $\Pi$. In addition to this, the APG produces a steeper logarithmic region, which is characterized by lower values of the von K\'arm\'an coefficient $\kappa$ and $B$, as well as decreased velocities in the buffer region. These effects, which were also observed by Monty {\it et al.} (2011) and Vinuesa {\it et al.} (2014), are due to the fact that the APG energizes the largest scales in the flow, which become shorter and more elongated, and have their footprint in their near-wall region. Also, these manifestations of the APG become more evident as $\beta$ increases. Moreover, comparisons of the Reynolds-stress tensor showed a progressive increase in the value of the inner peak of the streamwise turbulence intensity profile, as well as the development of an outer peak which in the strong APG case ($\beta \simeq 14.1$) exceeds the magnitude of the inner peak. Note that the development of a more energetic outer region with increasing $\beta$ is also observed in the wall-normal and spanwise fluctuation profiles, as well as in the Reynolds shear stress. Comparison of the TKE budgets also shows the different energy distribution across the boundary layer when an APG is present, with increased production and dissipation profiles throughout the whole boundary layer. The emergence of an incipient outer peak in the production profile is observed at $\beta \simeq 14.1$, phenomenon which was also reported by Sk\r{a}re and Krogstad (1994). The increased dissipation leads to larger values of the viscous diffusion and the velocity-pressure-gradient correlation near the wall in order to balance the budget.  

Analysis of the inner-scaled premultiplied spanwise spectra showed the presence of the inner spectral peak at around $y^{+}_{n} \simeq 12$ and $\lambda_{z}^{+} \simeq 120$, in agreement with the observations by Eitel-Amor {\it et al.} (2014) in ZPG TBLs at higher $Re_{\theta}$ up to $8,300$. As in the inner peak of $\overline{u^{2}_{t}}^{+}$, the spectral near-wall peak increases with the magnitude of the APG, as a consequence of the energizing process of the large structures in the flow, which have their footprint close to the wall. Also as a consequence of this energizing process, an outer spectral peak emerges at strong APGs with $\beta \simeq 4.1$, which is responsible for the development of larger outer region values in all the components of the Reynolds stress tensor. This spectral outer peak is observed at wavelengths of around $\lambda_{z} \simeq 0.65 \delta_{99}$, closer to the wall than the outer peak observed at $Re_{\theta} \simeq 8,300$ by Eitel-Amor {\it et al.} (2014)  in the ZPG case, with $\lambda_{z} \simeq 0.8 \delta_{99}$. At this point it is not possible to state whether this difference arises from low-$Re$ effects, or from a mechanism of energy transfer to the larger scales fundamentally different between high-$Re$ at ZPG and the effect of the APG. 

Future studies at higher Reynolds numbers will be aimed at further assessing the connections between the effect of APGs on the large-scale motions in the flow and the effect of $Re$ in ZPG boundary layers, in order to separate pressure-gradient and Reynolds-number effects.


\Acknowledgments

RV and PS acknowledge financial support from the Swedish Research Council (VR) and the Knut and Alice Wallenberg Foundation. The simulations were performed on resources provided by the Swedish National Infrastructure for Computing (SNIC) at the Center for Parallel Computers (PDC), in KTH, Stockholm. 

%
\begin{References}
\item Coles, D. (1956), The law of the wake in the turbulent boundary layer, {\it J. Fluid Mech.}, Vol. 1, pp. 191--226. 

\item Jones, L. E., Sandberg, R. D. and Sandham, N. D. (2008), Direct numerical simulations of forced and unforced separation bubbles on an airfoil at incidence, {\it J. Fluid Mech.}, Vol. 602, pp. 175--207. 

\item Alferez, N., Mary, I. and Lamballais, E. (2013), Study of stall development around an airfoil by means of high fidelity large eddy simulation, {\it Flow Turb. Comb.}, Vol. 91, pp. 623--641. 

\item Fischer, P. F., Lottes, J. W. and Kerkemeier, S. G. (2008), NEK5000: Open source spectral element CFD solver. Available at: \url{http://nek5000.mcs.anl.gov}. 

\item Schlatter, P. and \"Orl\"u, R. (2012), Turbulent boundary layers at moderate Reynolds numbers: inflow length and tripping effects, {\it J. Fluid Mech.}, Vol. 710, pp. 5--34.

\item Hosseini, S. M., Vinuesa, R., Schlatter, P., Hanifi, A. and Henningson, D. S. (2016), Direct numerical simulation of the flow around a wing section at moderate Reynolds number, {\it Int. J. Heat Fluid Flow}, doi:10.1016/j.ijheatfluidflow.2016.02.001

\item Monty, J., Harun, Z. and Marusic, I. (2011), A parametric study of adverse pressure gradient turbulent boundary layers, {\it Int. J. Heat Fluid Flow}, Vol. 32, pp. 575--585. 

\item Vinuesa, R., Rozier, P. H., Schlatter, P. and Nagib, H. M. (2014), Experiments and computations of localized pressure gradients with different history effects, {\it AIAA J.}, Vol. 52, pp. 368--384.

\item Schlatter, P. and \"Orl\"u, R. (2010), Assessment of direct numerical simulation data of turbulent boundary layers, {\it J. Fluid Mech.}, Vol. 659, pp. 116--126. 

\item Vinuesa, R., Bobke, A., \"Orl\"u, R. and Schlatter, P. (2016), On determining characteristic length scales in pressure-gradient turbulent boundary layers, {\it Phys. Fluids}, Vol. 28, pp. 055101.

\item Sk\r{a}re, P. E. and Krogstad, P.-\r{A}. (1994), A turbulent equilibrium boundary layer near separation, {\it J. Fluid Mech.}, Vol. 272, pp. 319--348. 

\item Harun, Z., Monty, J. P., Mathis, R. and Marusic, I. (2013), Pressure gradient effects on the large-scale structure of turbulent boundary layers, {\it J. Fuid Mech.}, Vol. 715, pp. 477--498.

\item Hutchins, N. and Marusic, I. (2007), Large-scale influences in near-wall turbulence, {\it Phil. Trans. R. Soc. A}, Vol. 365, pp. 647--664.

\item Vallikivi, M., Hultmark, M. and Smits, A. J.  (2015), Turbulent boundary layer statistics at very high Reynolds number, {\it J. Fluid Mech.}, Vol. 779, pp. 371--389.

\item Eitel-Amor, G., \"Orl\"u, R. and Schlatter, P.  (2014), Simulation and validation of a spatially evolving turbulent boundary layer up to $Re_{\theta}=8300$, {\it Int. J. Heat Fluid Flow}, Vol. 47, pp. 57--69.

\item Lin, J., Laval, J. P., Foucaut, J. M. and Stanislas, M. (2008), Quantitative characterization of coherent structures in the buffer layer of near-wall turbulence. Part 1: Streaks, {\it Exp. Fluids}, Vol. 45, pp. 999--1013.

\end{References}
\end{document}